\documentclass[fleqn,12pt,twoside]{article}
\usepackage[headings]{espcrc1}

\readRCS
$Id: espcrc1.tex,v 1.2 2004/02/24 11:22:11 spepping Exp $
\usepackage{graphicx}
\newcommand{\AmS}{{\protect\the\textfont2
  A\kern-.1667em\lower.5ex\hbox{M}\kern-.125emS}}

\title{Interferometry search for new forms of matter in A+A collisions}

\author{S.V. Akkelin\address[MCSD]{Bogolyubov Institute for Theoretical Physics, Kiev
03143, Metrologichna 14b, Ukraine},
       Yu.M. Sinyukov\addressmark}

\runtitle{Interferometry search for new forms of matter in A+A
collisions} \runauthor{S.V. Akkelin, Yu.M. Sinyukov}

\begin{document}

% typeset front matter
\maketitle

\begin{abstract}
A method allowing studies of the hadronic matter at the early
evolution stage in A+A collisions is developed. It is based on an
interferometry analysis of approximately conserved values such as
the averaged phase-space density (APSD) and the specific entropy of
thermal pions. The plateau found in the APSD behavior vs collision
energy at SPS is associated, apparently, with the deconfinement
phase transition at low SPS energies; a saturation of this quantity
at the RHIC energies indicates the limiting Hagedorn temperature for
hadronic matter.
\end{abstract}

\section{INTRODUCTION}

The goal of  experiments with heavy ion collisions is to study the
new forms of matter which can be created under an extreme
conditions at the early stage of the evolution. The bulk of
hadronic observables  are related, however, only to the very last
period of the matter evolution  - the end of the collective
expansion  when the system decays.  Our basic idea is to use the
''conserved observables'' which are integrals of motion (IM) to
study the properties of the matter at the hadronization stage at
different energies of A+A collisions. A structure of the IM,
besides the trivial ones, depends on a scenario of the matter
evolution.  Our analysis is based on hydrodynamic picture of the
chemically frozen evolution. Then the average phase-space density
(APSD) and specific entropy (SE) of pions as well as transverse
spectra slopes and interferometry volumes are approximately
conserved quantities \cite{AkkSin}. A conservation of the latter
two quantities was found also in transport models \cite{Amelin}
and for the final stage of the hydrodynamic evolution in Ref.
\cite{Csorgo}. The APSD and SE can be expressed through the
observed spectra and interferometry radii irrespectively the form
of freeze-out (isothermal) hypersurface and transverse flows on it
\cite{AkkSin,AkkSin1}. This makes it possible an analysis of the
hadronization stage.

\section{THE APSD AND ENTROPY AS OBSERVABLES IN A+A COLLISIONS}

The method is based on the standard approach for spectra formation
 that supposes that thermal freeze-out in expanding
 locally equilibrated system happens  at some space-time hypersurface
 with uniform temperature
$T$ and particle number density $n$. Then, within this
approximation which is probably appropriate in some
''boost-invariant'' mid-rapidity interval, $\Delta y < 1$, the
locally-equilibrium phase-space density of pions,
 totally averaged over the
hypersurface of thermal freeze-out, $\sigma =\sigma _{th}$, and
momentum except the longitudinal one (rapidity is fixed, e.g.,
$y=0$) is the same as the totally averaged phase-space density in
the static homogeneous Bose gas \cite{AkkSin} and can be extracted
directly from the experimental data in full accordance with the
pioneer Bertsch idea \cite{Bertsch}:
\begin{equation}
(2\pi )^{3}\left\langle f(\sigma ,y)\right\rangle _{y=0}=\frac{\int d^{3}p%
\overline{f}_{eq}^{2}}{\int d^{3}p\overline{f}_{eq}}=\kappa
\frac{2\pi ^{5/2}\int \left( \frac{1}{R_{O}R_{S}R_{L}}\left(
\frac{d^{2}N}{2\pi m_{T}dm_{T}dy}\right) ^{2}\right) dm_{T}}{dN/dy},
\label{result1}
\end{equation}
where $\overline{f}_{eq}\equiv (\exp (\beta (p_{0}-\mu )-1)^{-1}$,
and $\beta $ and $\mu $ coincide with the inverse of the
temperature and chemical potential at the freeze-out hypersurface.
The $\kappa =1$ if  one ignores the resonance decays. Using the
same approximation of the uniform freeze-out temperature and
density  we get the following expression for specific entropy in
mid-rapidity:
\begin{equation}
s\equiv\frac{dS/dy}{dN/dy}=\frac{\int d^{3}{p}\,[-\overline{f}_{eq}\ln \overline{f}%
_{eq}+(1+\overline{f}_{eq})\ln (1+\overline{f}_{eq})]}{\int d^{3}p\overline{f%
}_{eq}}.  \label{en-to-n}
\end{equation}
The above ratio  depends only on the two parameters: the
temperature and chemical potential at freeze-out. The temperature
can be obtained from the fit of the transverse spectra for
different particle species and we will use the value $T=120$ MeV
as a typical "average value" for SPS and RHIC experiments. Another
parameter, the chemical potential, we extract from an analysis of
the APSD following to (\ref{result1}).  The factor $\kappa $  is
accounting for a contribution of the short-lived resonances to the
spectra and interferometry radii and absorbs also the effect of
suppression of the correlation function due to the long-lived
resonances \cite{AkkSin}. Because of the chemical freeze-out a big
part of pions, about a half, are produced by the short-lived
resonances after thermal freeze-out.   To estimate the thermal
characteristics and "conserved observables" at the final stage of
hydrodynamic evolution by means of Eqs. (\ref{result1}),
(\ref{en-to-n}) one needs to eliminate non-thermal contributions
to the pion spectra and correlation functions from resonance
decays at post freeze-out stage.  To do this we use the results of
Ref. \cite{AkkSin} where a study of the corresponding
contributions within hydrodynamic approach gives the values of
parameter $\kappa$ to be $\kappa =0.65$ for SPS and $\kappa =0.7$
for RHIC, if half of pions is produced by the resonances at post
freeze-out stage. Then, from Eq. (\ref{result1}) one can extract
the pion chemical potential at thermal freeze-out. This makes it
possible to estimate the APSD, the specific entropy
 and other thermal parameters of the system at the end
of the hydrodynamic expansion.

\section{THE ANALYSIS OF EXPERIMENTAL DATA AND THE RESULTS}

To evaluate the APSD of negative  pions by means of Eq.
(\ref{result1}) we utilize the yields, transverse momentum spectra
and interferometry radii of $\pi ^{-}$ at mid-rapidity measured in
central heavy ions collisions by the  E895 and E802 Collaborations
for AGS energies \cite{E895,E802}, NA49 Collaboration for SPS CERN
energies \cite{NA49}, STAR and PHENIX Collaborations for RHIC BNL
energies \cite{STAR,PHENIX,PHENIX1}. Since the interferometry
radii are measured by PHENIX Collaboration for $0-30$ $\%$
centrality events at $\sqrt{s_{NN}}=200$ GeV \cite{PHENIX1}, we
increase the interferometry volume measured by PHENIX
Collaboration at this c.m. energy  by a factor of  $1.215$ to get
the interferometry volume corresponding to the most central $0-5$
$\%$ centrality bin in accordance with $N_{part}$ dependence of
the Bertsch-Pratt radius parameters found in Ref. \cite{PHENIX1}.
The results for the APSD at mid-rapidity for \textit{all} negative
pions ($\kappa =1$) at the AGS, SPS, RHIC energies  are presented
in Fig. 1. The calculated APSD are mostly influenced by the values
of the HBT radii in the low $p_{T}$ region and by the choice of
analytic
\begin{figure}[htb]
\begin{minipage}[t]{5.0cm}
\includegraphics*[scale=.25]{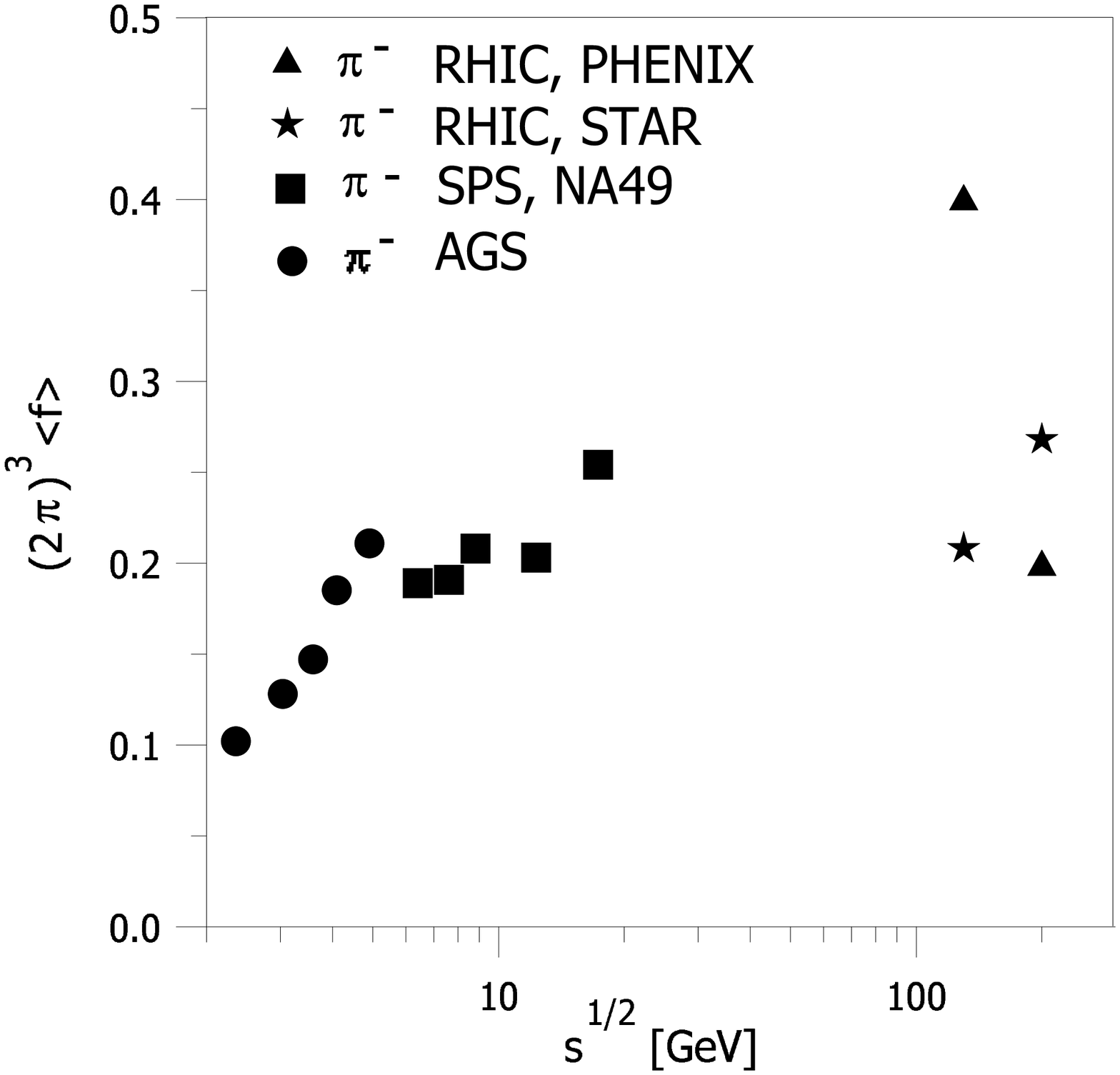}
\caption{The extracted from the experimental data
\cite{E895,E802,NA49,STAR,PHENIX,PHENIX1}
  APSD (see
 Eq. (\ref{result1})) of all negative pions at
mid-rapidity  as function of c.m. energy per nucleon in heavy ion
central collisions.} \label{fig:phtot}
\end{minipage}
\hspace{\fill}
\begin{minipage}[t]{5.0cm}
\includegraphics*[scale=.25]{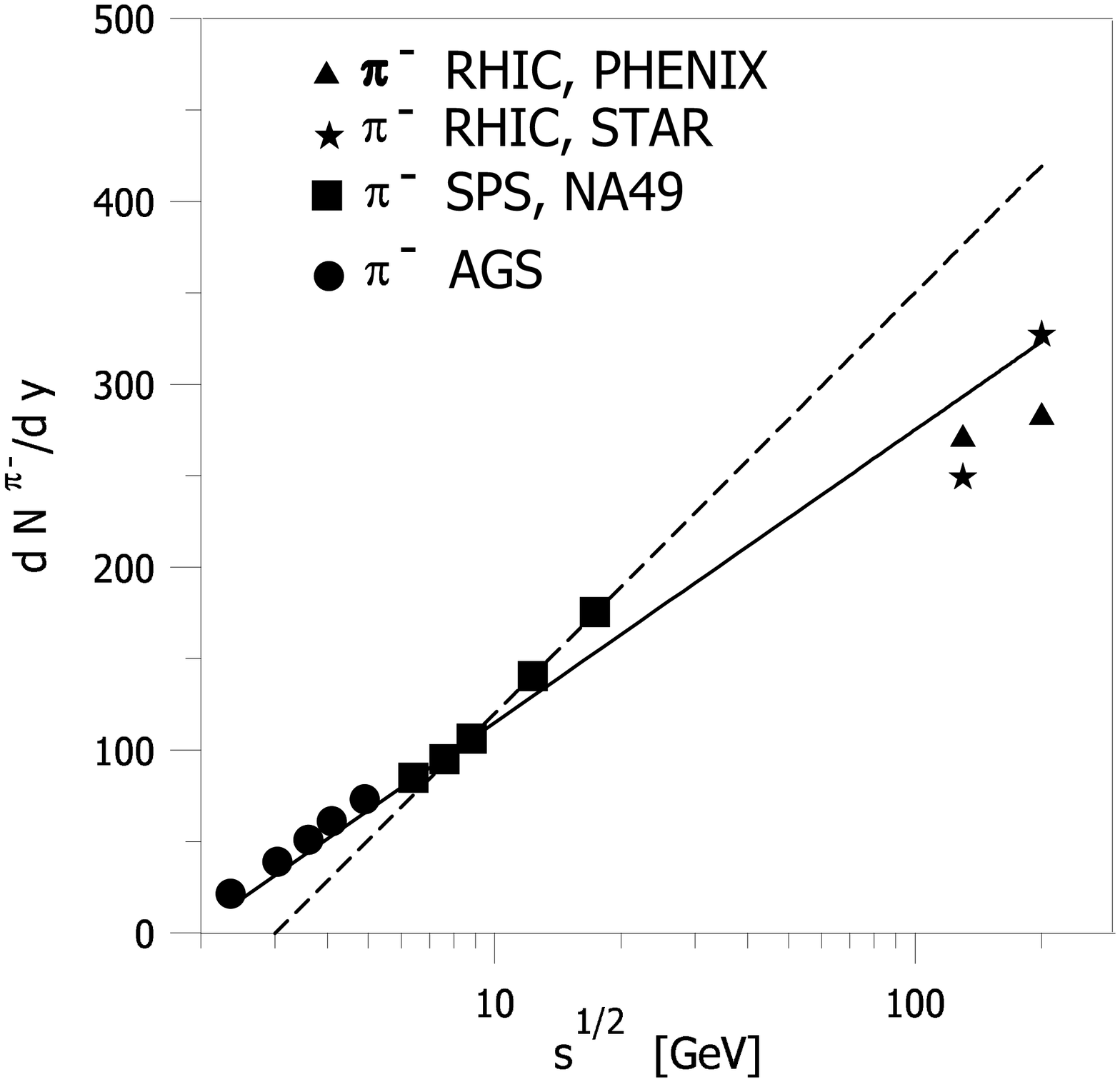}
 \caption{The  rapidity density of all negative pions \cite{E895,E802,NA49,STAR,PHENIX}
  as function of c.m. energy per nucleon in heavy
ion central collisions.} \label{fig:ntoen}
\end{minipage}
\hspace{\fill}
\begin{minipage}[t]{5.0cm}
\includegraphics*[scale=.25]{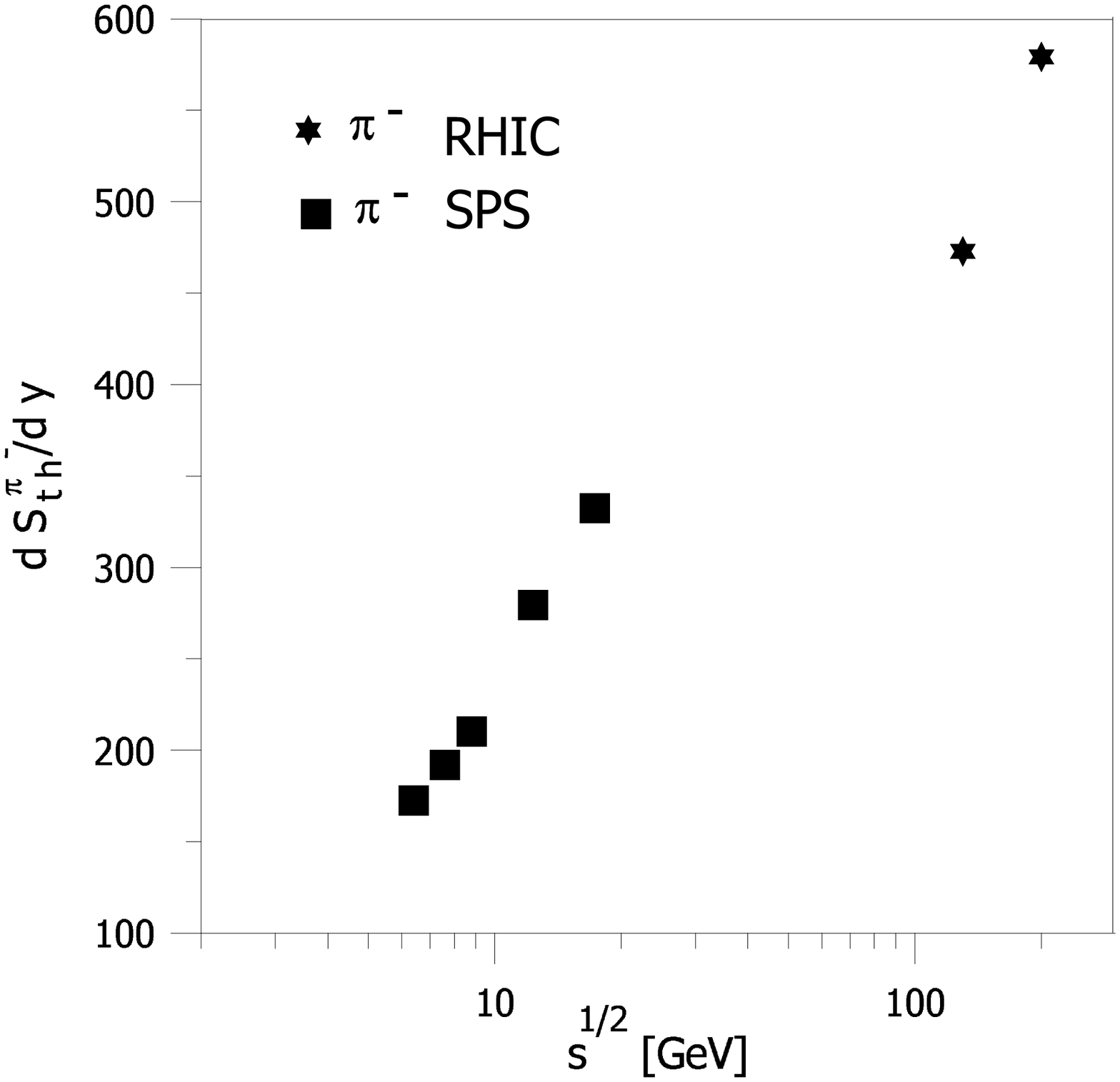}
\caption{The calculated following to the Eq. (\ref{en-to-n})
rapidity density of entropy for negative thermal pions  as
function of c.m. energy per nucleon in heavy ion central
collisions. } \label{fig:entrop}
\end{minipage}
\end{figure}
 $p_{T}$ parameterizations behind the lower measured
$p_{T}$ bin in accordance with the experimental data tendency, the
differences there can result in significant differences in the
calculated APSD values, as one can see from Fig. 1 for PHENIX vs
STAR data.
 Since the part of the experimental data that we used is
preliminary and was presented by the Collaborations without error
bars, we do not demonstrate error bars at the figures. The APSD of
negative \textit{thermal} pions are evaluated from the APSD of
\textit{all} negative  pions ($\kappa =0.65$ for SPS and $\kappa
=0.7$ for RHIC energies respectively) and are used then to extract
the pion chemical potentials $\mu$ at thermal freeze-out at
different SPS and RHIC energies, and after that to calculate the
specific entropies, $s_{th}^{\pi^{-}}$ (\ref{en-to-n}), and the
entropies,
$\frac{dS_{th}^{\pi^{-}}}{dy}=s_{th}^{\pi^{-}}\frac{dN_{th}^{\pi^{-}}}{dy}$,
of negative \textit{thermal} pions. In Fig. 2 we present the
rapidity densities, $dN^{\pi^{-}}/dy$, of \textit{all} negative
pions at mid-rapidity in central nucleus-nucleus collisions for
the AGS, SPS and RHIC energies. The lines in Fig. 2 represent the
logarithmic law of energy dependence for negative pion
multiplicities: $a\log_{10} (\sqrt{s_{NN}}/b)$, where $a=160
(230)$, $b = 1.91$ GeV (3 GeV) for solid (dashed) lines
respectively. The entropies $dS_{th}^{\pi^{-}}/dy$ of negative
\textit{thermal} pions are demonstrated in Fig. 3, where the
values at the RHIC energies are mean values of STAR and PHENIX
data.
\section{CONCLUSIONS}
 A behavior of the pion APSD vs
collision energy has a plateau at low SPS energies that indicates,
apparently, the transformation of initial energy to non-hadronic
forms of matter at SPS; a saturation of that quantity at the RHIC
energies can be treated as an existence of the limiting Hagedorn
temperature of hadronic matter, or maximal temperature of
deconfinement $T_c$. A behavior of the  entropy of thermal  pions
and measured pion multiplicities in central rapidity region vs
energy demonstrates an anomalously high slope of an increase of the
pion entropy/multiplicities at SPS energies compared to what takes
place at the AGS  and RHIC energies. This additional growth could
be, probably, a manifestation of the QCD critical end point (CEP).
The observed phenomenon can be caused by the dissipative effects
that usually accompany phase transitions, such as an increase of the
bulk viscosity \cite{Gyulassy}, and also by peculiarities of pionic
decays of $\sigma$ mesons and other resonances with masses that are
reduced, as compare to its vacuum values, in vicinity of the QCD CEP
\cite{Stephanov}. At the RHIC energies there is no anomalous rise of
pion entropy/multiplicities, apparently, because the crossover
transition takes place far from the CEP and no additional degrees of
freedom appear at that scale of energies: quarks and gluons were
liberated at previous energy scale.

\section*{ACKNOWLEDGMENTS}

 The research described in this publication was
made possible in part  by Award No. UKP1-2613-KV-04 of the U.S.
Civilian Research $\&$ Development Foundation for the Independent
States of the Former Soviet Union (CRDF), by NATO Collaborative
Linkage Grant No. PST.CLG.980086 and the ERG (GDRE): Heavy ions at
ultrarelativistic energies - a European Research Group.

\end{document}